\documentclass[twocolumn,aps,showpacs,floatfix,superscriptaddress]{revtex4}
\pdfoutput=1 
\usepackage{amsmath,amssymb,eucal,graphicx,float,epstopdf}

\begin{document}
\title{Irreversible Reactions and Diffusive Escape: Stationary Properties}

\author{P.~L.~Krapivsky}
\affiliation{Department of Physics, Boston University, Boston,
Massachusetts 02215}
\author{E.~Ben-Naim}
\affiliation{Theoretical Division and Center for Nonlinear Studies,
Los Alamos National Laboratory, Los Alamos, New Mexico 87545}

\begin{abstract}
We study three basic diffusion-controlled reaction
processes---annihilation, coalescence, and aggregation. We examine the
evolution starting with the most natural inhomogeneous initial
configuration where a half-line is uniformly filled by particles,
while the complementary half-line is empty. We show that the total number of particles that infiltrate the initially empty 
half-line is finite and has a stationary distribution.  We determine the evolution of the
average density from which we derive the average
total number $N$ of particles in the initially empty half-line; e.g.,
for annihilation $\langle N\rangle = \frac{3}{16}+\frac{1}{4\pi}$. For
the coalescence process, we devise a procedure that in principle
allows one to compute $P(N)$, the probability to find exactly $N$
particles in the initially empty half-line; we complete the
calculations in the first non-trivial case ($N=1$).  As a by-product
we derive the distance distribution between the two leading particles.
\end{abstract}


\maketitle

\section{Introduction}

\begin{figure}
\centering
\includegraphics[width=7.6cm]{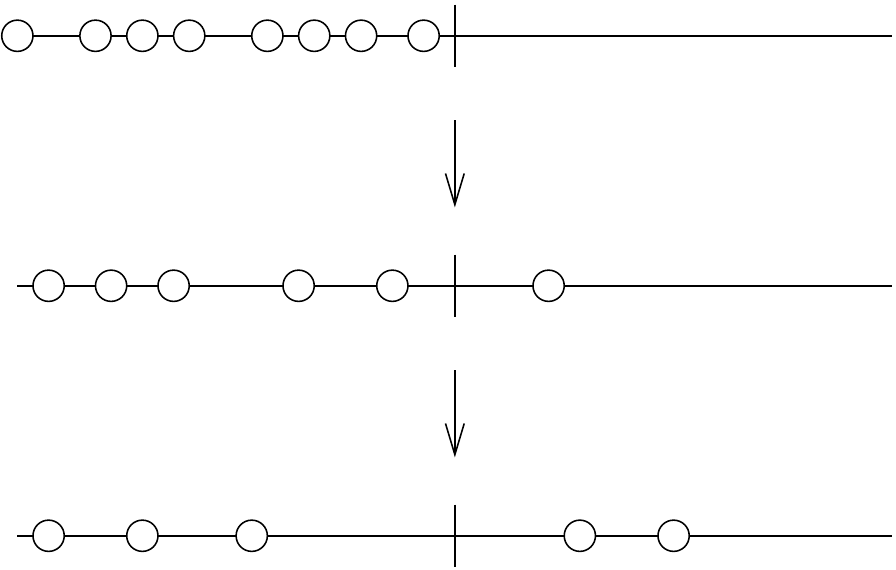}
\caption{Illustration of the irreversible reaction process. The 
initial condition with fixed concentration of particles left of the
origin and no particles to the right of the origin is shown on the top. The next two layers illustrate 
the state of the system at later times (the latest time at the bottom) with a dwindling
concentration of particles left of the origin and a fluctuating {\em
finite} total number of particles to the right of the origin. }
\label{Fig:ill}
\end{figure}

Reaction-diffusion systems are ubiquitous in biology, chemistry and physics. An important class of such systems, diffusion-controlled processes in which a reaction happens whenever two reactants ``meet'', has been studied for almost a century (see
e.g. \cite{SM17,RZ17,C43,agg-rev,smoke77,BG80,A81,OTB89,cloud,ABD90,thv,AH00,book}). Mathematically,
these processes are strongly interacting infinite particle systems, so
it is not surprising that many basic questions remain unanswered.  Here we
analyze a few basic diffusion-controlled processes focusing on the
interplay between reaction and spatial inhomogeneity of the initial
setting. We will see that for an extreme inhomogeneous initial configuration with an empty half-line (\ref{Fig:ill}), the statistics of the occupation number of this half-line is asymptotically stationary and highly non-trivial.

We study three diffusion-controlled reaction systems. One is a single-species annihilation which is represented by the reaction scheme
\begin{equation}
\label{annih}
A+A\to \emptyset
\end{equation}
Identical particles (denoted by $A$) undergo diffusion and whenever two particles touch each other, they disappear. Strict annihilation is rare. One example is the annihilation of domain walls in an
Ising spin chain subjected to a zero-temperature spin-flip
dynamics \cite{book}. In most situations, however, annihilation merely implies
that the reaction product does not further affect the reaction
process. For instance, a collision between two atoms may lead to the
formation of a diatomic molecule; if such molecules are stable (that
is, they do not break back into atoms) and if they do not influence
diffusing atoms, we can use the
reaction scheme \eqref{annih}.

Another simple diffusion-controlled process, coalescence, can be
represented by the reaction scheme
\begin{equation}
\label{coal}
A+A\to A
\end{equation}
Strict coalescence is also rare. It occurs for example in a Potts
chain with infinitely many states evolving according to the
non-conserved zero-temperature dynamics \cite{book}. In most
situations, the coalescence process is a coarse-grained description of
the aggregation process
\begin{equation}
\label{aggr}
A_i+A_j\to A_{i+j}
\end{equation}
where $A_m$ represents clusters of mass $m$, i.e., clusters composed
of $m$ monomers. The symbolic representation \eqref{coal} indicates that
we disregard the mass of aggregates and hence it faithfully describes
the gross features of the aggregation process when the diffusion
constant is mass-independent and when the size is also
mass-independent.

Little is known about evolution of irreversible reactions starting
from spatially inhomogeneous configurations. Here we consider the
simplest situation when at time $t=0$ the half space $x<0$ is
uniformly filled by particles, while the half space $x>0$ is empty
(see Fig.~\ref{Fig:ill})
\begin{equation}
\label{step}
c(x,t=0) =
\begin{cases}
\rho       & x<0\\
0           & x>0
\end{cases}
\end{equation}
This step-function initial condition has been used e.g. to study front propagation in reaction-diffusion processes \cite{bd,mvs} and shock waves in driven exclusion processes \cite{dls}, it is also popular in experimental \cite{exp:cold1,exp:cold2} and theoretical (see e.g. \cite{theory:cold1,theory:cold2,theory:cold3,theory:cold4}) studies of cold quantum gases, and in analyses of quantum spin chains (\cite{chain1,chain2,chain3,chain4} and references therein).

In three dimensions, one can use the rate equation approach to analyze
the evolution. For annihilation and coalescence processes, the density $c(x,t)$ satisfies the 
reaction-diffusion equation, $\partial_t c = D\partial_{xx} c - Kc^2$, where $D$ is the diffusion coefficient and $K$ the reaction rate. The long time behavior is simple, viz. in the scaling limit
\begin{equation}
\label{xtX}
|x|\to \infty, \quad  t\to\infty, \quad  X = \frac{x}{\sqrt{2Dt}}=\text{finite},
\end{equation}
the density becomes $c(x,t) = (Kt)^{-1}C(X)$. The scaled density
$C(X)$ satisfies an ordinary differential equation which can be solved
numerically.

In this paper we focus on the one-dimensional case.  This is the most challenging setting and the emergent results are richer than in higher dimensions. The challenge stems from the fact that in one dimension one cannot employ the rate equation approach, i.e., one cannot rely on reaction-diffusion equations. The behaviors are
still well understood when the initial conditions are homogeneous,
e.g., $c\sim (Dt)^{-1/2}$, see e.g. \cite{BG80,A81,AH00,book}. For the inhomogeneous initial condition (Fig.~\ref{Fig:ill}) the density eventually decays in the same manner everywhere. However, particles infiltrate only up to distance $\sim \sqrt{Dt}$ into the vacant half-line, so the total {\em number} of particles in the initially vacant half-line is expected to be finite and governed by a stationary distribution. In this paper, we establish this behavior analytically.

For the inhomogeneous setting \eqref{step}, the scaling
form
\begin{equation}
\label{cC:1d}
c(x,t) = \frac{1}{\sqrt{2\pi D t}}\,C(X)
\end{equation}
is anticipated. We shall obtain the scaled density profile 
$C(X)$ analytically. An immediate consequence of \eqref{cC:1d} is that the average
total number of particles $\langle N\rangle$ in the initially empty
half-line is {\em finite}. More precisely, for the coalescence process
\begin{equation}
\label{Nav:c}
\langle N\rangle_\text{c} = \frac{3}{8}+\frac{1}{2\pi} =   0.534154943\ldots
\end{equation}
For the annihilation process, the value is twice smaller. Thus the total number of particles $N$ in the initially empty
half-line is a fluctuating quantity with a finite average. (The total number of particles in the initially filled
half-line is of course infinite, although the density vanishes algebraically with time $\sim
(Dt)^{-1/2}$.) Furthermore, we also find that the total number of
particles $N$ is governed by a stationary probability distribution
$P(N)$.

The determination $P_\text{c}(N)$ for the coalescence process is
technically challenging. In addition to
$P_\text{c}(0)=\frac{1}{2}$ which can be established on the basis of
symmetry alone, we are able to compute the probability
\begin{eqnarray}
\label{CP:one}
P_\text{c}(1) &=& \frac{11\pi -4}{16\pi}
+ \frac{1}{2\pi}\!\left[\arctan\!\left(\tfrac{1}{\sqrt{8}}\right)-2\arctan\!\left(\tfrac{1}{\sqrt{2}}\right) \right] \nonumber\\
                     &=& 0.466095976\ldots
\end{eqnarray}
to have exactly one particle in the initially empty half-line. With
much more effort it could be possible to determine $P_\text{c}(2)$,
but to find the entire distribution $P_\text{c}(N)$ requires more
advanced methods.

For the annihilation process we haven't been able even to determine
$P_\text{a}(0)$. Indeed, the coalescence process is more
tractable than the annihilation process. This subtle difference
between these two very similar diffusion-controlled reaction processes
was already encountered in the homogeneous setting. For instance, for
the coalescence process the distance distribution between adjacent
particles is well-known (see \cite{AH00,book}), while for the
annihilation process the distance distribution has not been determined
analytically \cite{AbA95,DZ96,KB97}. Another long-standing unresolved
problem is to determine the long-time behavior of an impurity (a
particle with diffusion coefficient generally different from the
diffusion coefficient of the host particles) in the annihilation
process \cite{KBR94,CM96}.

The annihilation process with the step-function initial condition \eqref{step} was investigated 
in Ref.~\cite{LPS98}. That study was focused on the survival
probabilities $S_n(t)$, where $n=1,2,\ldots$ labels particles
according to their initial locations: $x_1(0) > x_2(0) > x_3(0)$ etc. These survival probabilities exhibit an
intriguing temporal behavior, the decay laws depend only on the parity of the
original label: $S_n\sim t^{-\alpha}$ for even $n$ and as $S_n\sim
t^{-\beta}$ for odd $n$; the decay exponents $\alpha$ and $\beta$ are
still unknown \cite{LPS98}.

The rest of this article is organized as follows. In Sec.~\ref{sec:1d}
we consider diffusion-controlled annihilation, coalescence, and
aggregation processes in one dimension and compute the average
densities everywhere.  Statistics of the total number of particles in
the initially empty half-line is studied in Sec.~\ref{sec:empty}. In Sec.~\ref{sec:concl} we summarize our results and discuss several open problems. 

\section{Density Profile} 
\label{sec:1d}

In one dimension, we compute the density profile using the empty
interval method which allows an efficient analytical treatment of the
one-dimensional diffusion-controlled coalescence process
(Sec.~\ref{DCP}). Modifications of the empty interval method which are
suitable for annihilation and aggregation are presented in
Sec.~\ref{DAP} and Sec.~\ref{sec:AP}.

We consider continuous versions of the aggregation, annihilation and
coalescence processes in which size-less particles undergo Brownian
motions on a line. Lattice versions in which each particle occupies
a site on the one-dimensional lattice and hops to nearest-neighboring
sites are also tractable. The basic asymptotic behaviors are 
identical in both versions.

\subsection{Coalescence Process}
\label{DCP}

In the one-dimensional diffusion-controlled coalescence process, point
particles undergo identical independent Brownian motions and merge
instantaneously whenever they meet. An elegant exact treatment of this
strongly interacting infinite particle system is possible through the
empty interval technique \cite{ABD90,AH00,book,AbA95,T89,AA93,MbA01}.

To explain the empty interval technique, we begin with spatially-homogeneous coalescence process. Let $E(x,y;t)$ be the probability that the interval $[x,y]$ is empty at time $t$. In the homogeneous case these probabilities
depend only on the length of the interval: $E(x,y;t)=E(\ell,t)$ where
$\ell=y-x$. The probabilities $E(\ell,t)$ satisfy the diffusion equation 
\begin{equation}
\label{Ezt}
\frac{\partial E(\ell,t)}{\partial t} = 2D\,\frac{\partial^2 E(\ell,t)}{\partial \ell^2}
\end{equation}
and the boundary condition
\begin{equation}
\label{BC:Ez}
E(\ell=0,t)=1
\end{equation}
The initial condition is $E(\ell,t=0)=e^{-\rho\ell}$ if the initial
distribution of particles is random and the initial density is
$\rho$. Note also the general formula for the concentration
\begin{equation}
\label{cE}
c(t) = - \frac{\partial E(\ell,t)}{\partial \ell}\Big|_{\ell=0}
\end{equation}

The physically relevant region is $\ell\geq 0$, but we can extend
\eqref{Ezt} to $\ell<0$. It proves convenient to make a shift,
$F(\ell,t) = E(\ell,t) - 1$, so that the boundary condition
\eqref{BC:Ez} becomes $F(\ell=0,t) = 0$.  The governing equation for
$F(\ell,t)$ remains the diffusion equation
\begin{equation}
\label{Fzt}
\frac{\partial F}{\partial t} = 2D\,\frac{\partial^2 F}{\partial \ell^2}
\end{equation}
We have $F(\ell,t=0)= e^{-\rho\ell}-1$ when $\ell>0$ and we seek the
initial condition for $\ell<0$ in such a way that the boundary
condition $F(\ell=0,t) = 0$ is manifestly obeyed. The proper choice is
\begin{equation}
\label{IC:Fz}
F(\ell,t=0)=
\begin{cases}
\exp[-\rho\ell] - 1 &\ell>0\\
1- \exp[\rho\ell]   &\ell<0
\end{cases}
\end{equation}
One can write an exact solution of the diffusion equation \eqref{Fzt}
subject to the initial condition \eqref{IC:Fz}. We do not display this
solution since our main goal is to establish the long-time behavior,
and for that purpose, we can replace \eqref{IC:Fz} by the simplified
initial condition
\begin{equation}
\label{IC:Fsimple}
F(\ell,t=0)=
\begin{cases}
- 1 &\ell>0\\
1   &\ell<0
\end{cases}
\end{equation}
The solution to \eqref{Fzt} and \eqref{IC:Fsimple} is $F=-\text{Erf}(\ell/\sqrt{8Dt})$, and therefore
\begin{equation}
\label{Ezt:sol}
E(\ell,t) = \text{Erfc}\left(\frac{\ell}{\sqrt{8Dt}}\right)
\end{equation}
By combining \eqref{cE} and \eqref{Ezt:sol} we recover the long-time asymptotic behavior of the density 
\begin{equation}
\label{c_uniform}
c(t) = \frac{1}{\sqrt{2\pi D t}}
\end{equation}

\begin{figure}
\centering
\includegraphics[width=8cm]{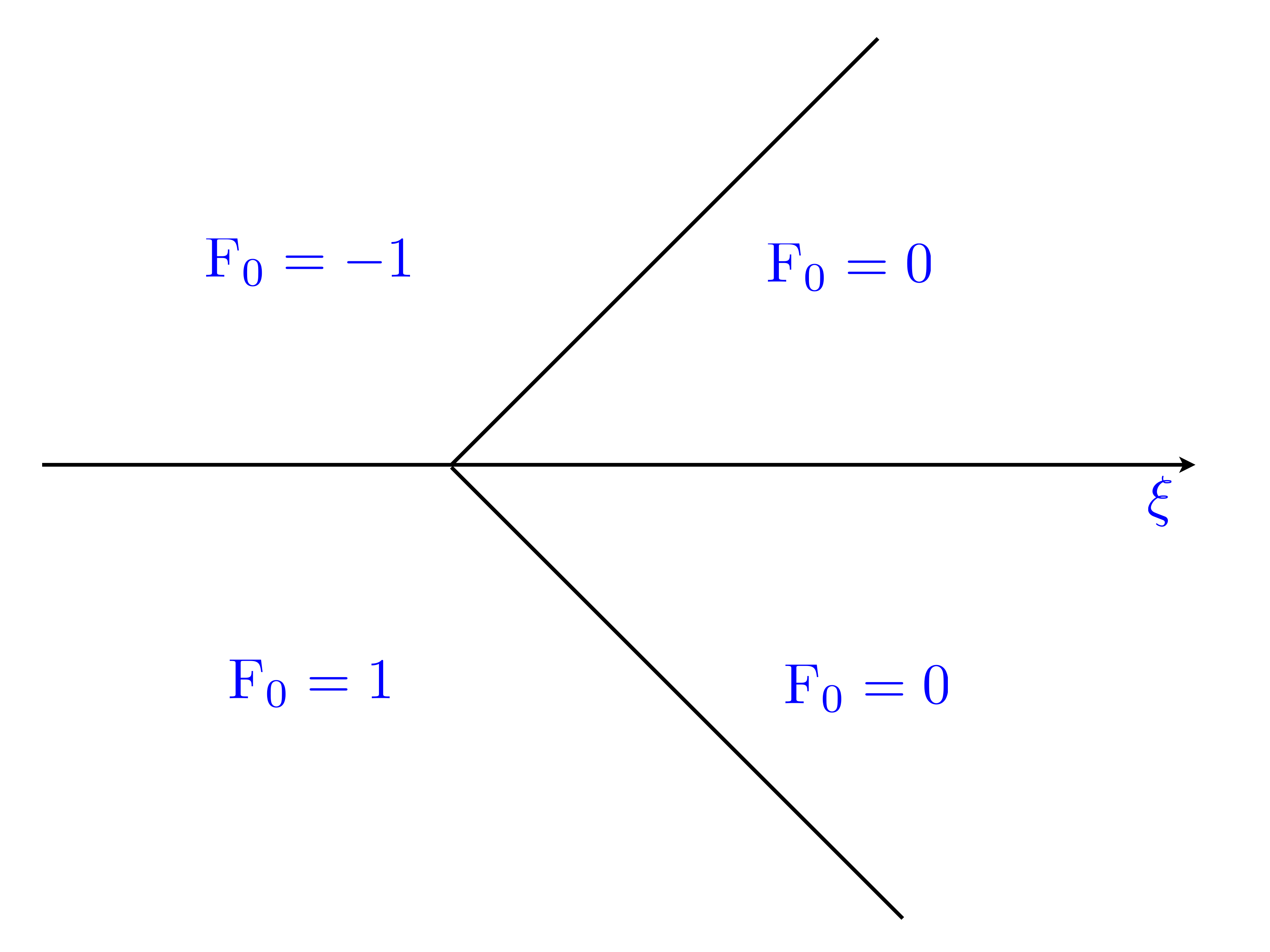}
\caption{The initial condition $F_0(\xi, \eta)=F(\xi, \eta; t=0)$ in
the $(\xi, \eta)$ plane. The vanishing of $F(\xi, \eta; t)$ on the
$\xi$ axis, $F|_{\eta=0}=0$, is manifest. }
\label{Fig:IC_F}
\end{figure}

We now turn to the spatially-inhomogeneous initial condition
\eqref{step}. The mathematical formulation is a natural generalization
of Eqs.~\eqref{Ezt}--\eqref{cE}. The governing equations are
\begin{equation}
\label{Exyt}
\frac{\partial }{\partial t}\,E(x,y; t) = D\left(\frac{\partial^2 }{\partial x^2}+\frac{\partial^2 }{\partial y^2}\right)E(x,y; t)
\end{equation}
The boundary condition becomes
\begin{equation}
\label{BC:Exy}
E(x,x;t)=1
\end{equation}
The density is found from
\begin{equation}
\label{cxt}
c(x,t) = - \frac{\partial E(x,y; t)}{\partial y}\Big|_{y=x}
\end{equation}
We have to solve \eqref{Exyt} subject to the initial condition
\begin{equation}
\label{IC:Exy}
E_0(x,y)=
\begin{cases}
1                       &0 < x < y\\
\exp[\rho x]       &x < 0 < y\\
\exp[\rho(x-y)]   &x < y <0 
\end{cases}
\end{equation}
Writing $E(x,y; t) = 1 + F(x,y; t)$ and focusing on the long-time limit we can again use a simplified initial condition:
\begin{equation}
\label{IC:Fxy}
F_0(x,y)=
\begin{cases}
0                 &0 <  \text{min}(x, y)\\
-1                &x < 0, ~x < y\\
1                 &y <0, ~y< x 
\end{cases}
\end{equation}
Changing variables
\begin{equation}
\xi = y+x, \quad \eta = y-x
\end{equation}
we end up with the two-dimensional diffusion equation
\begin{equation}
\label{Fxyt}
\frac{\partial }{\partial t}\,F(\xi,\eta; t) = 2D\left(\frac{\partial^2 }{\partial \xi^2}+\frac{\partial^2 }{\partial \eta^2}\right)
F(\xi, \eta; t)
\end{equation}
subject to the initial condition [Fig.~\ref{Fig:IC_F}]
\begin{equation}
\label{IC:F}
F_0(\xi, \eta)=
\begin{cases}
0                 &0 <  \xi, ~~ -\xi<\eta<\xi\\
-1                &0<\eta,  ~~ \xi < \eta\\
1                 &\eta <0, ~~ \xi < -\eta
\end{cases}
\end{equation}
Expressing the boundary condition \eqref{BC:Exy} in terms of the
auxiliary function $F(\xi, \eta; t)$ gives $F(\xi, \eta=0; t)=0$. This
boundary condition is indeed manifestly obeyed with the choice of
initial condition \eqref{IC:F}, see Fig.~\ref{Fig:IC_F}.

The solution to equation \eqref{Fxyt}--\eqref{IC:F} is
straightforward. One gets
\begin{eqnarray}
\label{Exyt:sol}
E(x,y; t) & = & 1 -  \tfrac{1}{2} \text{Erfc}\left(\tfrac{X+Y}{2}\right)\text{Erf}\left(\tfrac{Y-X}{2}\right) \nonumber\\
&-&\frac{1}{\pi}\int_{-\frac{X+Y}{2}}^\infty du\,e^{-u^2}\int_{u+X}^{u+Y} dv\,e^{-v^2}
\end{eqnarray}
where $X=x/\sqrt{2Dt}$ and $Y=y/\sqrt{2Dt}$. By substituting \eqref{Exyt:sol} into \eqref{cxt} we confirm the scaling form \eqref{cC:1d} and obtain the density profile 
\begin{equation}
\label{CX:1d}
C(X) = \frac{1}{2} \text{Erfc}(X) +  \frac{1}{\sqrt{8}}\,e^{-X^2/2}\,\text{Erfc}\!\left(-\frac{X}{\sqrt{2}}\right)
\end{equation}
The density profile \eqref{CX:1d} holds everywhere and it does {\em not} depend on the initial density $\rho$. The latter property is known in the homogeneous case, so it must hold when $X\to -\infty$; remarkably, it remains true even for the initially empty half-line. The dependence on the initial density can be observed e.g. for $x\sim t$, but for any fixed $X$ this dependence disappears in the long time limit. 

\begin{figure}
\centering
\includegraphics[width=8cm]{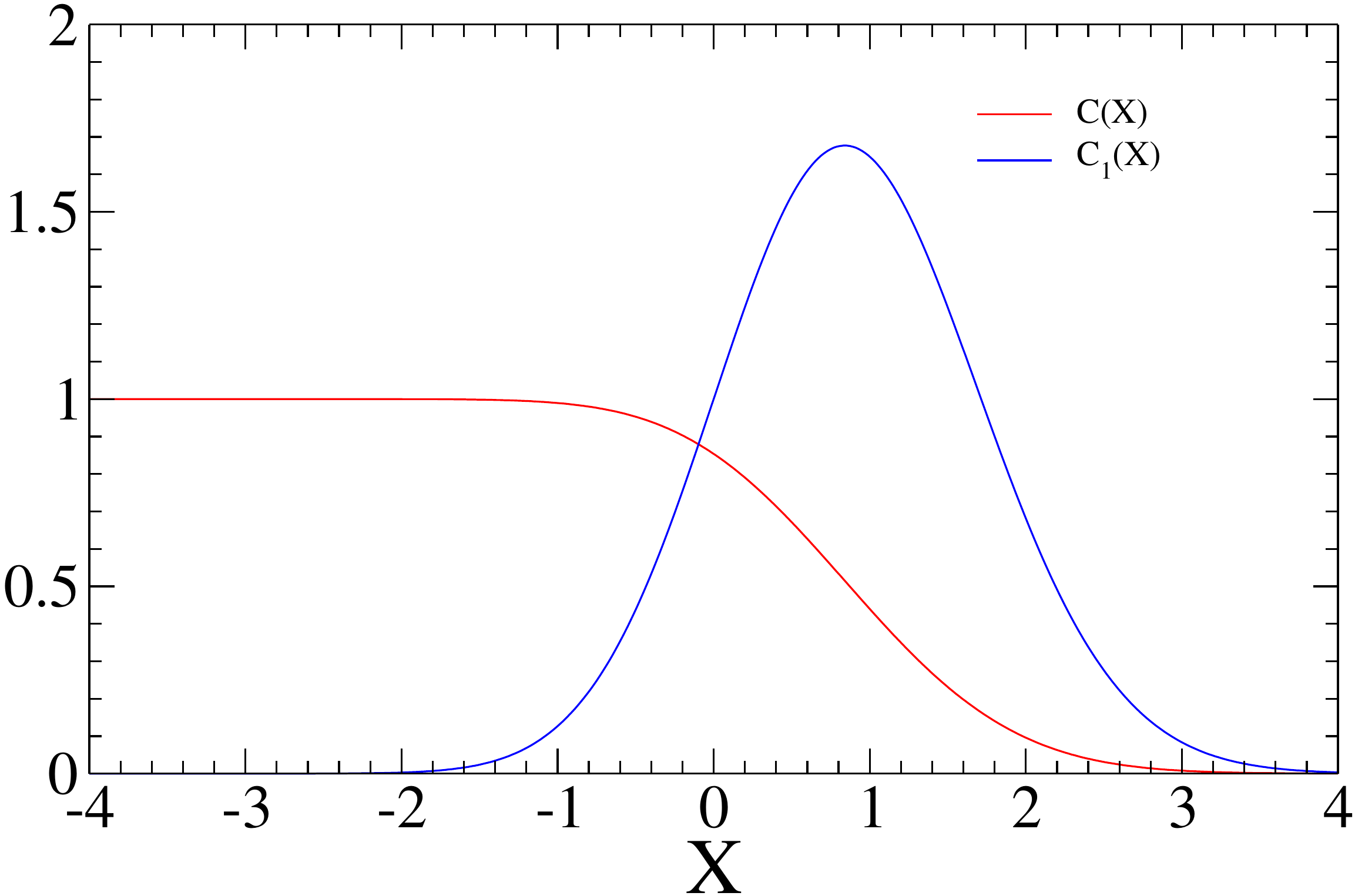}
\caption{The scaled density $C(X)$ in all three processes and the
scaled density of monomers $C_1(X)$ for the aggregation process. The
density $C(X)$ is the same, Eq.~\eqref{CX:1d}, both for the
coalescence and annihilation processes if the scaling forms are chosen
according to \eqref{cC:1d} for the coalescence process and
\eqref{cCA:1d} for the annihilation process. In the aggregation
process, the scaled densities of clusters of small mass are given by
the same Eq.~\eqref{CmX} as the density of monomers.}
\label{Fig:CXnew}
\end{figure}

On the interface separating the two half-lines, we have
$C(X=0)=\frac{1}{2}+\frac{1}{\sqrt{8}} = 0.853553\ldots$ (see
Fig.~\ref{Fig:CXnew}). If there were no reactions, only diffusion,
$c(x,t) = \frac{1}{2} \text{Erfc}\!\left(\frac{X}{\sqrt{2}}\right)$,
so on the interface separating the two half spaces the density is
exactly a half that of the bulk.

The exact solution \eqref{Exyt:sol} contains information beyond the
density. For instance, specializing \eqref{Exyt:sol} to $Y=\infty$
gives the probability $\mathcal{E}(x; t)\equiv E(x, y=\infty; t)$ that
the half-line $(x,\infty)$ is empty. This probability is a function
of a single scaled variable:
\begin{equation*}
\mathcal{E}(x; t)=\mathcal{E}(X) = 1-\frac{1}{2\sqrt{\pi}}\int_{-\infty}^\infty du\,e^{-u^2}\,\text{Erfc}(u+X)
\end{equation*}
The integral can be expressed through the error function:
\begin{equation}
\label{EX}
\mathcal{E}(X) = 1-\frac{1}{2}\,\text{Erfc}\left(\frac{X}{\sqrt{2}}\right)
\end{equation}
The behavior $\mathcal{E}(0) = \frac{1}{2}$ simply reflects that the
right-most particle in the coalescence process is equally likely to be
to the left and to the right of the origin [see also
\eqref{coal:zero}].

\subsection{Irreversible Annihilation Process}
\label{DAP}

For the irreversible annihilation process \eqref{annih}, instead of empty interval
probabilities one considers $G(x,y; t)$, the probability that there is
an even number of particles in the interval $[x, y]$. The
probabilities $G(x,y; t)$ satisfy the same equations
\eqref{Exyt}--\eqref{cxt} as the empty interval probabilities (see
\cite{AbA95,MbA01}), the only difference is the initial condition,
viz. instead of \eqref{IC:Exy} one gets
\begin{equation*}
\label{IC:Gxy}
G_0(x,y)=\frac{1}{2}+\frac{1}{2}\times 
\begin{cases}
1                          &0 < x < y\\
\exp[2\rho x]        &x < 0 < y\\
\exp[2\rho(x-y)]   &x < y <0 
\end{cases}
\end{equation*}
There is therefore a duality between annihilation and coalescence, namely
\begin{equation}
G(x,y; t)=\tfrac{1}{2}+\tfrac{1}{2}E(x,y; t)
\end{equation}
This duality holds if the initial densities differ by factor 2
exactly:
\begin{equation}
\rho^\text{annihilation} = \tfrac{1}{2} \rho^\text{coalescence}
\end{equation}
This remarkable duality between annihilation and coalescence extends
to more complicated correlation functions. It was found
\cite{HOS95,KPWH95,BRD95,S95,MbA01} in the context of homogeneous
setting, and it also holds in our spatially-inhomogeneous
setting.

In the long-time limit, the initial densities are irrelevant, to
leading order, and therefore, we have
\begin{eqnarray}
\label{Gxyt:sol}
G(x,y; t) & = & 1 -  \tfrac{1}{4} \text{Erfc}\left(\tfrac{X+Y}{2}\right)\text{Erf}\left(\tfrac{Y-X}{2}\right) \nonumber\\
&-&\frac{1}{2\pi}\int_{-\frac{X+Y}{2}}^\infty du\,e^{-u^2}\int_{u+X}^{u+Y} dv\,e^{-v^2}
\end{eqnarray}
The density has the scaling form
\begin{equation}
\label{cCA:1d}
c_\text{a}(x,t) = \frac{1}{\sqrt{8\pi D t}}\,C(X)
\end{equation}
with the same scaled density $C(X)$ as before, Eq.~\eqref{CX:1d}.

Specializing \eqref{Gxyt:sol} to $Y=\infty$ gives the probability
$\mathcal{E}_\text{a}(x; t)\equiv G(x, y=\infty; t)$ that the
half-line $(x,\infty)$ contains an even number of particles:
\begin{equation}
\label{even_X}
\mathcal{E}_\text{a}(X) = 1-\frac{1}{4\sqrt{\pi}}\int_{-\infty}^\infty du\,e^{-u^2}\,\text{Erfc}(u+X)
\end{equation}
In particular, the initially-empty half line contains an even number
of particles with probability
\begin{equation}
\label{even_a}
\mathcal{E}_\text{a}(0) = \sum_{N\geq 0}P_\text{a}(2N) = \frac{3}{4}
\end{equation}
 
\subsection{Irreversible Aggregation Process}
\label{sec:AP}

The average mass density $\mu(x,t)=\sum_{m\geq 1}
mc_m(x,t)$ is not affected by the aggregation process and hence it
satisfies the diffusion equation $\partial_t \mu = D\partial_{xx} \mu$. Therefore
\begin{equation}
\label{mass:sol}
\mu(x,t) = \frac{\rho}{2}\,\text{Erfc}\left(\frac{x}{\sqrt{4Dt}}\right)= \frac{\rho}{2}\,\text{Erfc}\left(\frac{X}{\sqrt{2}}\right)
\end{equation}

To probe other average characteristics, we need more advanced
methods. It turns out \cite{T89,AA93} that instead of empty interval
probabilities it suffices to use $P_m(x,y;t)$, the probability that
the total mass inside the interval $[x,y]$ equals $m$. Assuming that
all clusters are point particles that diffuse with the same diffusion
constant $D$, one finds that the probabilities $P_m(x,y;t)$ satisfy
\begin{equation}
\label{Pxyt}
\frac{\partial }{\partial t}\,P_m(x,y; t) = D\left(\frac{\partial^2 }{\partial x^2}+\frac{\partial^2 }{\partial y^2}\right)P_m(x,y; t)
\end{equation}
The boundary condition \eqref{BC:Exy} generalizes to 
\begin{equation}
\label{BC:Pxy}
P_m(x,x;t)=\delta_{m,0}
\end{equation}
The mass distribution is found from
\begin{equation}
\label{cmxt:def}
c_m(x,t) =  \frac{\partial P_m(x,y; t)}{\partial y}\Big|_{y=x}
\end{equation}
The initial condition reads
\begin{equation}
\label{IC:Pxy}
P_m(t=0)=
\begin{cases}
\delta_{m,0}                                         &0 < x < y\\
\frac{(\rho|x|)^m}{m!}\,e^{\rho x}          &x < 0 < y\\
\frac{[\rho(y-x)]^m}{m!}e^{\rho(x-y)}    &x < y <0 
\end{cases}
\end{equation}

The governing equations \eqref{Pxyt} are linear, yet their analysis is rather cumbersome, see Appendix~\ref{app:aggr-1d}. Here we just present the major findings. In the scaling limit \eqref{xtX}, the mass distribution acquires the scaling form
\begin{equation}
\label{cmXt:scaling}
c_m(X,t) = \frac{1}{4\pi \rho Dt}\,C_m(X)
\end{equation}
For $m=O(1)$, the scaled densities become mass independent, i.e., the
same as the monomer density
\begin{equation}
\label{CmX}
C_m(X) = \sqrt{\frac{\pi}{2}}\,X e^{-X^2/2}\,\text{Erfc}\left(-\tfrac{X}{\sqrt{2}}\right)+e^{-X^2}
\end{equation}
As a function of the scaled distance, the monomer density has a
single peak (Fig.~\ref{Fig:CXnew}) and it is maximal at
$X\approx 0.84$.

One anticipates that the mass density $C_m(X)$ depends on the scaled mass:
\begin{equation}
\label{CmX:scaling}
C_m(X) = \Phi(M,X), \qquad M = \frac{m}{\rho\sqrt{2\pi Dt}}
\end{equation}

We now use the sum rule $\sum_{m\geq 1} c_m(X,t)=c(X,t)$, the scaled
mass distribution \eqref{cmXt:scaling}, \eqref{CmX:scaling} together
with the cluster density \eqref{cC:1d} to obtain an integral relation
for the scaled mass density
\begin{equation}
\label{IR:1}
\int_0^\infty dM\,\Phi(M,X)\! =\! \text{Erfc}(X) +  \frac{e^{-X^2/2}}{\sqrt{2}}\,\text{Erfc}\!\left(-\tfrac{X}{\sqrt{2}}\right)
\end{equation}
Similarly the sum rule $\sum_{m\geq 1} mc_m(X,t)=\mu(X,t)$ together with the exact result \eqref{mass:sol} for the mass density lead to another integral relation
\begin{equation}
\label{IR:2}
\int_0^\infty dM\,M\Phi(M,X)  = \text{Erfc}\!\left(\tfrac{X}{\sqrt{2}}\right)      
\end{equation}
The small mass tail of $\Phi(M,X)$  is given by \eqref{CmX}; it would be interesting to derive the large mass tail. 

\section{Particle Number}
\label{sec:empty}

Next, we study the total number of
particles $N(t)$ that infiltrate the initially empty half-line. This natural
quantity is {\em finite}, and it fluctuates throughout the
evolution. Hence, the average does not fully characterize the
probability distribution function. The finiteness of $N(t)$ follows
from a heuristic argument. The decay law $c\sim
(Dt)^{-1/2}$ in the homogeneous case and the fact that particles infiltrate by distance of the order of $\sqrt{Dt}$ into initially empty half-line
suggest that $N$ is indeed of the order of one. We can 
compute the average total number of particles $\langle N\rangle$ in
the initially empty half-line. For the coalescence process, we use
\eqref{cC:1d} to obtain
\begin{equation}
\label{Nav:int}
\langle N\rangle_\text{c} = \int_0^\infty dx\,c(x,t) = \frac{1}{\sqrt{\pi}}\int_0^\infty dX\,C(X)
\end{equation}
which in conjunction with the density profile \eqref{CX:1d} leads to the announced
result \eqref{Nav:c}. For the annihilation process, the density is exactly
two times smaller; accordingly, the average total number of particles
is 
\begin{equation}
\label{Nav:a}
\langle N\rangle_\text{a} = \frac{3}{16}+\frac{1}{4\pi} = 0.267077471\ldots
\end{equation}

To learn more about the random quantity $N$, one needs to compute its
probability distribution $P(N)$. The distribution functions for the
reaction processes \eqref{annih} and \eqref{coal} are actually
different. We are interested in both $P_\text{c}(N)$ and
$P_\text{a}(N)$. (We distinguish the two cases with subscripts.)  The
coalescence process is simpler, and a few exact results are possible.

The simplest exact result is the probability of finding no particles
in the initially empty half-line:
\begin{equation}
\label{coal:zero}
P_\text{c}(0)=\frac{1}{2}
\end{equation}
To derive Eq.~\eqref{coal:zero} consider the right-most
particle $R$ and notice that from its `view-point' the coalescence
process \eqref{coal} reduces to $A+R\to R$. Thus the right-most
particle is not affected by the coalescence process, so it performs a one-dimensional
Brownian motion. With probability $\frac{1}{2}$ the right-most is at $x<0$, and this is equivalent to saying that $N=0$. 

The normalization requirement, $\sum_{N\geq 0}P_\text{c}(N)=1$,
together with \eqref{coal:zero} indicates that $\sum_{N\geq
1}P_\text{c}(N)=\frac{1}{2}$. Combining this result with $\langle
N\rangle_\text{c} = \sum_{N\geq 1}NP_\text{c}(N)$ and
Eq.~\eqref{Nav:c} we obtain the sum rule
\begin{equation}
\label{Nc:sum}
\sum_{N\geq 2}(N-1)P_\text{c}(N) = \frac{1}{2\pi} - \frac{1}{8} = 0.0341549431\ldots
\end{equation}
This sum rule leads to the upper bound for the probability to find two particles: 
\begin{equation}
\label{coal2:bound}
P_\text{c}(2)<0.0341549431\ldots
\end{equation}
 
Combining the normalization requirement with \eqref{coal:zero} and \eqref{CP:one} we similarly derive the sum rule
\begin{eqnarray*}
\sum_{N\geq 3}(N-2)P_\text{c}(N) &=& 
\frac{1}{16}+\frac{1}{4\pi}\\
&+&\frac{1}{2\pi}\!\left[\arctan\!\left(\tfrac{1}{\sqrt{8}}\right)-2\arctan\!\left(\tfrac{1}{\sqrt{2}}\right) \right] \\
& = & 0.00025091951\ldots
\end{eqnarray*}
which yields the upper bound for the probability to find three particles in the initially empty half-line:
\begin{equation}
\label{coal3:bound}
P_\text{c}(3)<0.00025091951\ldots
\end{equation}

\subsection{Computation of $P_\text{c}(1)$}
\label{sec:P1}

We now consider the coalescence process and show that the probability
$P_\text{c}(1)$ to have exactly one particle in the initially empty
half-line is given by \eqref{CP:one}. In principle, our procedure can
be extended to $P_\text{c}(N)$ with arbitrary $N$; yet the computation
quickly becomes prohibitive. 

To determine $P_\text{c}(1)$ we use probabilities for two intervals to
be empty. In Sec.~\ref{DCP} we computed the probability $E(x,y)$ that
the interval $(x,y)$ is empty at time $t$. (In the following we do not
display the time variable, so $E(x,y)$ denotes $E(x,y| t)$, etc.) Let
$E(x_1,y_1; x_2, y_2)$ be the probability that the intervals
$(x_1,y_1)$ and $(x_2, y_2)$ are empty. We shall assume that
$x_1<y_1<x_2<y_2$, so the intervals are non-overlapping. The
probability $E(1;2)\equiv E(x_1,y_1; x_2, y_2)$ satisfies
\begin{equation*}
\frac{\partial }{\partial t}\,E(1; 2) = D\left(\frac{\partial^2 }{\partial x_1^2}+\frac{\partial^2 }{\partial y_1^2} +\frac{\partial^2 }{\partial x_2^2}+\frac{\partial^2 }{\partial y_2^2}\right)E(1; 2)
\end{equation*}
The solution of this equation, with obvious boundary conditions
\begin{equation*}
\begin{split}
& \lim_{x_1\uparrow y_1}E(x_1,y_1; x_2, y_2) = E(x_2, y_2)\\
& \lim_{y_1\uparrow x_2}E(x_1,y_1; x_2, y_2) = E(x_1, y_2)\\
& \lim_{x_2\uparrow y_2}E(x_1,y_1; x_2, y_2) = E(x_1, y_1)
\end{split}
\end{equation*}
can be expressed through the single-interval empty probabilities \cite{MbA01}:
\begin{eqnarray}
\label{E2}
E(1; 2) &=& E(x_1,y_1)E(x_2, y_2)-E(x_1,x_2)E(y_1,y_2)\nonumber\\
                                   & + & E(x_1,y_2)E(y_1,x_2)
\end{eqnarray}

For our purposes, it suffices to consider a simple subset of empty
interval probabilities, namely those with $y_2=\infty$. We denote
$\mathcal{E}(x,y,z)=E(x,y; z, \infty)$, in analogy with notation
$\mathcal{E}(z)=E(z, \infty)$ which we used in
Sec.~\ref{DCP}. Specifying \eqref{E2} to this setting we get
\begin{equation}
\label{E21}
\mathcal{E}(x,y,z) = E(x,y)\mathcal{E}(z)-E(x,z)\mathcal{E}(y) +  \mathcal{E}(x)E(y,z)
\end{equation}
Applying $\frac{\partial^2}{\partial x \partial z}$ to Eq.~\eqref{E21}
and taking the $z\to y$ limit, we find the probability $R(x,y)$
that the two right-most particles are at $x$ and $y$:
\begin{eqnarray}
\label{Rxy}
R(x,y) &=& \frac{\partial  E(x,y)}{\partial x}\, \frac{\partial \mathcal{E}(y)}{\partial y}
-\mathcal{E}(y)\,\frac{\partial^2  E(x,y)}{\partial x \partial y}\nonumber\\
 & - & c(y)\,\frac{\partial \mathcal{E}(x)}{\partial x}
\end{eqnarray}
Further, the probability $P_\text{c}(1)$ to have exactly one particle
in the initially empty half-line can be obtained by integrating the
probability density $R(x,y)$:
\begin{equation}
\label{PR}
P_\text{c}(1) = \int_0^\infty dy \int_{-\infty}^0 dx\,R(x,y)
\end{equation}

Computation of the integrals in Eq.~\eqref{PR}, detailed in Appendix~\ref{app:integral}, leads to the the announced expression \eqref{CP:one}. 

As a by-product of these calculations, we can determine the probability density $R(x,y)$
that the first and the second right-most particles are located at $x$
and $y$. Using \eqref{Rxy} we can express $R(x,y)$ in the scaling form
\begin{subequations}
\begin{align}
\label{Rxyt}
R(x,y| t) &= \frac{1}{\pi \sqrt{32}}\, (Dt)^{-1}\,\mathcal{R}(X,Y) \\
\label{RXY}
\mathcal{R}(X,Y) &= e^{-(Y-X)^2/4}\,\text{Erfc}\left(\tfrac{X+Y}{2}\right)\Psi(X,Y) \nonumber \\
&-e^{-X^2/2}\,\text{Erfc}(Y)
\end{align}
\end{subequations}
where we have used the shorthand notation
\begin{equation*}
\Psi(X,Y) =\ e^{-Y^2/2}+
(Y-X)\sqrt{2\pi}\left[1-\tfrac{1}{2}\text{Erfc}\left(\tfrac{Y}{\sqrt{2}}\right)\right]
\end{equation*}

For the homogeneous coalescence process, the distribution of distance between adjacent particles is easy to determine, see e.g. \cite{AH00,book}. For the particles on the edge, however, the
computation is involved and the scaled distance distribution
\eqref{RXY} apparently has not been known. For the annihilation
process, even in the bulk (equivalently for the homogeneous setting)
the distance distribution has not been established despite of the
considerable effort \cite{AbA95,DZ96,KB97}.

\subsection{Simulation Results}

We performed numerical simulations to measure various statistical
properties including in particular, those of the total number of
particle $N$.  In the simulations, we considered the lattice version
of the diffusion-controlled coalesce and annihilation
processes. Namely, we assumed that particles undergo (continuous time)
random walk on the one-dimensional lattice, and instantaneously
coalesce (or annihilate) whenever two particles occupy the same
site. In the long-time limit, the continuous (with particles undergoing
Brownian motion) and the lattice versions lead to the same results.

The initial condition become irrelevant in the long time limit, so we
considered the simplest initial condition when half of the lattice is fully
occupied. Thus initially every lattice site in the half-line $x<0$ is
occupied and every lattice site in the half-line $x\geq 0$ is empty.

\begin{table}[h]
\begin{tabular}{|c|l|l|}
\hline 
$N$ & $P_\text{a}(N)$ &  $P_\text{c}(N)$  \\
\hline
~0~ & ~0.74       & ~0.50\\
~1~ & ~0.25       & ~0.46 \\
~2~ & ~0.008     & ~0.03 \\
~3~ & ~$3\cdot 10^{-5}$ & ~$3\cdot 10^{-4}$\\
\hline
\end{tabular}
\caption{The probabilities $P_\text{a}(N)$ and $P_\text{c}(N)$ for $N\leq 3$ as obtained from numerical simulations.}
\end{table}

\begin{figure}
\centering
\includegraphics[width=8cm]{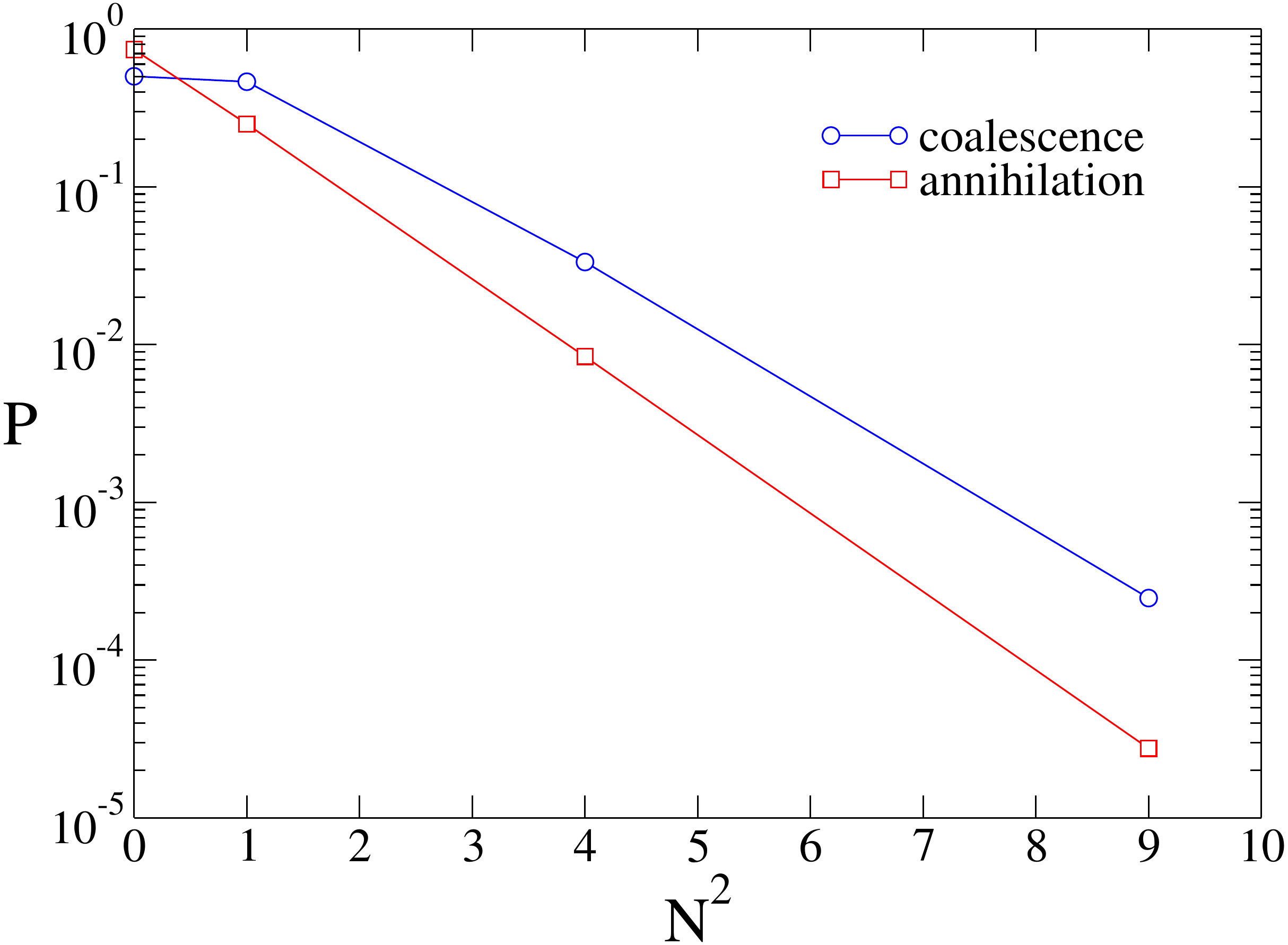}
\caption{Semi-logarithmic plots of the distributions $P_a(N)$ and $P_c(N)$ versus $N^2$.}
\label{Fig:pn}
\end{figure}

We measured the stationary distributions $P_a(N)$ and $P_c(N)$ for the number of
particles in the initially-empty half space ($x>0$) in the
annihilation and coalescence processes. The results are listed in
Table I. The numerical findings are consistent with the sum rule
\eqref{even_a}, viz. $\sum_{k\geq 0}P_\text{a}(2k)=\frac{3}{4}$.  For
the coalescence process, simulation results for $P_\text{c}(0)$ and
$P_\text{c}(1)$ are in excellent agreement with theoretical
predictions \eqref{coal:zero} and \eqref{CP:one}. The simulation
result for $P_\text{c}(2)$ is close to the theoretical upper bound
\eqref{coal2:bound} which is natural since $P_\text{c}(3)$ is very
small according to another theoretical upper bound
\eqref{coal3:bound}. Both distributions $P_a(N)$ and $P_c(N)$
apparently have Gaussian tails.  The numerical evidence (see Fig.~\ref{Fig:pn})
looks fairly convincing, even though the range in $N$ is very small.

\medskip

\subsection{Characteristics of the leader}

Here we outline what we know about the position of the ``leader'' defined as the
right-most particle. Features of interest include its average location
$\langle \ell\rangle$, the probability density $\Pi(\ell,t)$ that the
leader is at position $\ell$ at time $t$, etc. For the coalescence
process
\begin{equation}
\label{CP:extreme}
\langle \ell\rangle_\text{c} = 0, \quad  \Pi_\text{c}(\ell,t) = \frac{1}{\sqrt{4\pi D t}}\,\exp\!\left[-\frac{\ell^2}{4Dt}\right]
\end{equation}
For the annihilation process, in contrast, $\langle
\ell\rangle_\text{a}$ and $\Pi_\text{a}(\ell,t)$ are not known. Since
the initially empty half-line is empty with probability exceeding
$\frac{1}{2}$, one anticipates that the average
position of the leading particle recedes diffusively into the
initially occupied half-line. This was indeed observed in simulations
\cite{LPS98} where the probability density $\Pi_\text{a}(\ell,t)$ was
also measured, and found to be asymmetric and clearly non-Gaussian
\cite{LPS98}.

We studied numerically the one-dimensional aggregation process
\eqref{aggr} with mass-independent diffusion coefficients. We observed
that the average mass of the leader (the right-most aggregate) grows
as $\langle m\rangle = B \rho \sqrt{Dt}$ with $B=1.6 \pm 0.1$. This
growth is anticipated, e.g., using \eqref{mass:sol} one finds that the
average total mass in the initially empty half-line grows as
$\rho\sqrt{Dt/(4\pi)}$.

The one-dimensional aggregation process generalizes the coalescence
process, so it is more tractable than the annihilation process. The
properties of the leader might be amenable to exact analysis. We
already know the probability density $\Pi_\text{c}(\ell,t)$ for its
location. It will be interesting to determine the mass distribution
$P(m,t)$ and the more detailed joint distribution $\Pi(\ell,m,t)$,
i.e., the probability density that the leading particle has mass $m$
and located at $\ell$.

\section{Discussion}
\label{sec:concl}

In summary, we studied three basic one-dimensional diffusion-controlled
reaction processes---annihilation, coalescence, and aggregation. In
the case of aggregation, we assumed that the diffusion coefficients
are mass-independent, so it reduces to the coalescence process if one
focuses only on the total cluster distribution. Our theoretical analysis generalizes the empty interval method to
inhomogeneous initial conditions.

We examined the evolution starting with the inhomogeneous initial
configuration when a half-line is uniformly filled by particles, while
the complementary half-line is empty. We computed the average particle
density as a function of position. Our main finding is that while the
overall density profile is a time-dependent quantity, statistical
properties of the total number of particles residing in the
initially-empty half-space become independent of time, in the
long-time limit.  In particular, the total number of particles in the
initially empty half-line is finite and governed by a stationary
probability distribution.  We were able to compute analytically
several features of this distribution function, e.g. the average.

There are numerous basic questions about the behavior of
one-dimensional diffusion-controlled annihilation, coalescence, and
aggregation that remain unanswered. We limited ourselves to equal-time correlation functions. Many of them, e.g. $\langle N(t)\rangle$, become time-independent in the long time limit. Two-time correlation functions, such as  $\langle N(t_1) N(t_2)\rangle$, are natural extensions of our study. 

Another interesting time-dependent quantities are various first-passage (or persistence) characteristics \cite{Sid:FP,Bray13}. For instance, the survival probabilities of the particles in the annihilation process decay algebraically with time, and the decay exponent depends only on the parity of the label corresponding to the initial order; these two exponents \cite{LPS98} haven't been determined analytically. 

One can also ask about the probability $S(t)$ that not a single particle ever entered
the initially empty half-line during the time interval $(0,t)$. For
the coalescence process $S_\text{c}(t)\sim t^{-1/2}$ since the problem
reduces to the survival probability for a Brownian
particle in one dimension \cite{Sid:FP,Bray13}. For the
annihilation process $S_\text{a}(t)\sim t^{-3/16}$ as follows from
equivalence with spin persistence problem \cite{D95}. The probability of never entering the initially empty half-line is equivalent to the probability that the position of the leader satisfies $\ell(t')<0$ for all $t'<t$. Since the typical position of the leader scales diffusively, it is natural to ask about the probability that $\ell(t')<A\sqrt{D t'}$ for all $t'<t$, where $A$ is some fixed constant. This probability decays algebraically, $S(t,A)\sim t^{-\theta(A)}$ as $t\to \infty$. For the coalescence process we have effectively a single particle problem which is solvable and the exponent $\theta_\text{c}(A)$ is known \cite{KR96} for arbitrary $A$; for the annihilation process, only $\theta_\text{a}(0)=\frac{3}{16}$ is known \cite{D95}. Another natural generalization involves first-passage properties involving
the number of particles residing in the initially-empty half space \cite{bk-fp}. 

The lack of analytical tools for probing the properties of the annihilation process is frustrating, although not surprising in the light of earlier work \cite{AbA95,DZ96,KB97,LPS98,KBR94,CM96}. We think that establishing analytical tools for probing the annihilation process represent the most important and most challenging extension of the current work. Some delicate properties of the homogenous annihilation process have been derived analytically
\cite{D95} using field-theoretical methods; perhaps, the techniques of \cite{D95} can be extended to
the inhomogeneous setting.

\smallskip\noindent
We acknowledge support from US-DOE grant DE-AC52-06NA25396 (EB).

\appendix
\section{Scaling Behavior in the One-Dimensional Aggregation Process}
\label{app:aggr-1d}

One can try to solve \eqref{Pxyt} using the generating function technique. Here we use another approach relying on the recurrent nature of Eqs.~\eqref{Pxyt}. We first show in detail how to find the density of monomers and then generalize. The probability $P_1(x,y;t)$ that the total mass contained in the interval $[x,y]$ is equal to one is clearly the probability that the interval $[x,y]$ contains one monomer. This probability varies according to diffusion equation
\begin{equation}
\label{P1t}
\frac{\partial }{\partial t}\,P_1 = D\left(\frac{\partial^2 }{\partial x^2}+\frac{\partial^2 }{\partial y^2}\right)P_1
\end{equation}
supplemented by the boundary condition
\begin{equation}
\label{BC:P1}
P_1(x,x;t) = 0
\end{equation}
As earlier, we extend the definition of $P_1(x,y;t)$  from the physically relevant half-plane $x\leq y$ to the entire plane. The initial condition
\begin{equation}
\label{IC:P1}
P_1(t=0)=
\begin{cases}
0                                      &x>0, ~y>0\\
(-\rho x)\,e^{\rho x}          &x < 0 < y\\
\rho y \,e^{\rho y}            &y < 0 < x\\
\rho(y-x)\,e^{\rho(x-y)}    &x < y <0 \\
-\rho(x-y)\,e^{\rho(y-x)}   &y < x < 0 
\end{cases}
\end{equation}
agrees with \eqref{IC:Pxy} when $x\leq y$, while in the supplementary half-plane $x\geq y$ the choice made in Eq.~\eqref{IC:P1} assures that the boundary condition \eqref{BC:P1} is manifestly obeyed. 

Solving the diffusion equation \eqref{P1t} subject to the initial condition \eqref{IC:P1} yields
\begin{eqnarray*}
\frac{4\pi D t}{\rho} P_1 &=& \int_{-\infty}^0 dx_0 \int_0^\infty dy_0\, (-x_0)\,e^{\rho x_0} G\\
&+&\int_0^\infty dx_0 \int_{-\infty}^0 dy_0\, y_0\,e^{\rho y_0} G\\
&+&\int_{-\infty}^0 dy_0 \int_{-\infty}^{y_0} dx_0\,  (y_0-x_0)\,e^{\rho (x_0-y_0)} G\\
&+&\int_{-\infty}^0 dx_0 \int_{-\infty}^{x_0} dy_0\,  (y_0-x_0)\,e^{-\rho (x_0-y_0)} G
\end{eqnarray*}
Here $G=G(x,y,t| x_0, y_0)$ is the diffusion propagator:
\begin{equation*}
G(x,y,t| x_0, y_0) = \exp\!\left\{-\frac{(x-x_0)^2+(y-y_0)^2}{4Dt}\right\}
\end{equation*}

Recalling \eqref{cmxt:def} and performing straightforward calculations one finds the monomer density $c_1(x,t)$ 
\begin{eqnarray}
\label{c1xt}
\frac{8\pi (D t)^2}{\rho} c_1& = & \int_0^\infty dy_0\! \int_0^\infty dw\,w^2 e^{-\rho w} H_1 \\
&+& \int_0^\infty dy_0\int_0^\infty dx_0\, x_0 e^{-\rho x_0}(y_0-x_0) H_2 \nonumber
\end{eqnarray}
where we have used shorthand notation:
\begin{equation*}
\begin{split}
& H_1 = H(x| -y_0, -y_0-w)\\
& H_2 = H(x| -x_0, y_0)\\
&H(x| x_0, y_0) = \exp\!\left\{-\frac{(x-x_0)^2+(x-y_0)^2}{4Dt}\right\}
\end{split}
\end{equation*}

The integral representation \eqref{c1xt} is exact, that is, valid at all $t>0$. In the long-time limit, the monomer density simplifies. Let us first compute $c_1(0,t)$, the monomer density exactly on the interface. When $x=0$, the right-hand side of \eqref{c1xt} becomes
\begin{eqnarray*}
& & \int_0^\infty dy_0\! \int_0^\infty dw\,w^2 e^{-\rho w} \exp\!\left\{-\frac{y_0^2+(y_0+w)^2}{4Dt}\right\} \\
&& +\int_0^\infty dy_0\int_0^\infty dx_0\, x_0 e^{-\rho x_0}(y_0-x_0) \exp\!\left\{-\frac{x_0^2 + y_0^2}{4Dt}\right\}
\end{eqnarray*}
In the long-time limit, more precisely when $\rho^2 Dt\gg 1$, the integral in the second line dominates and asymptotically it grows as $2Dt/\rho^2$. Therefore 
\begin{equation}
c_1(0,t) = \frac{1}{4\pi \rho Dt}
\end{equation}
More generally, $c_1(x,t) = (4\pi \rho Dt)^{-1/2}$ for $x\sim \rho^{-1}$. 

In the scaling limit \eqref{xtX} with $X$ being fixed, the integral in the first line on the right-hand side of \eqref{c1xt} approaches to $\rho^{-3}(2\pi Dt)^{1/2}\,\text{Erfc}(X)$, while the integral in the second line of \eqref{c1xt} tends to 
\begin{equation*}
\label{2nd}
\frac{2Dt}{\rho^2}\left[\sqrt{\frac{\pi}{2}}\,X e^{-X^2/2}\,\text{Erfc}\left(-\tfrac{X}{\sqrt{2}}\right)+e^{-X^2}\right]
\end{equation*}
Thus
\begin{eqnarray}
\label{c1Xt:sol}
c_1(X,t) &=& \frac{1}{4\pi \rho Dt}
\left[\sqrt{\frac{\pi}{2}}\,X e^{-X^2/2}\,\text{Erfc}\left(-\tfrac{X}{\sqrt{2}}\right)+e^{-X^2}\right]\nonumber\\
&+& \frac{1}{2\rho^2}\,\frac{1}{\sqrt{8\pi (Dt)^3}}\,\text{Erfc}(X)
\end{eqnarray}
The term in the second line is negligible for finite $X$, both positive and negative, but we kept this term as it prevails when $X\ll -\sqrt{\ln(\rho^2 Dt)}$. This term determines the asymptotic monomer density far away from the interface (equivalently, for the uniform initial distribution): 
\begin{equation}
c_1(-\infty,t)= (8\pi)^{-1/2}\,\rho^{-2}\, (Dt)^{-3/2} 
\end{equation}

A similar treatment leads to
\begin{eqnarray*}
\label{cmxt}
\frac{8\pi (D t)^2}{\rho^m} c_m & = & 
\int_0^\infty dy_0\! \int_0^\infty dw\,\frac{w^{m+1}}{m!}\, e^{-\rho w} H_1 \\
&+& \int_0^\infty dx_0 \int_0^\infty dy_0\, \frac{x_0^m}{m!}\, e^{-\rho x_0}(y_0-x_0) H_2
\end{eqnarray*}
For clusters of finite mass, $m=O(1)$, one gets
\begin{eqnarray}
\label{cmXt:sol}
c_m(X,t) &=& \frac{1}{4\pi \rho Dt}
\left[\sqrt{\frac{\pi}{2}}\,X e^{-X^2/2}\,\text{Erfc}\left(-\tfrac{X}{\sqrt{2}}\right)+e^{-X^2}\right]\nonumber\\
&+& \frac{m+1}{4\rho^2}\,\frac{1}{\sqrt{8\pi (Dt)^3}}\,\text{Erfc}(X)
\end{eqnarray}
The term in the first line dominates for finite $X$, and it does {\em not} depend on the mass in the leading order. 

\section{Computation of the integrals in \eqref{PR}}
\label{app:integral}

Plugging \eqref{Rxy} into \eqref{PR} and performing the integration over $x$ we get
\begin{eqnarray}
\label{Pc1}
P_\text{c}(1) &=&  \int_0^\infty dy\, \frac{\partial \mathcal{E}(y)}{\partial y}\,[E(0,y)-E(-\infty,y)] \nonumber\\
&-&\int_0^\infty dy\, \mathcal{E}(y)\,\frac{\partial }{\partial y}\,[E(0,y)-E(-\infty,y)] \nonumber\\
&-&\int_0^\infty dy\, c(y)[\mathcal{E}(0)-\mathcal{E}(-\infty)]
\end{eqnarray}
Since $E(-\infty,y)=\mathcal{E}(-\infty)=0$ and $\mathcal{E}(0)=\frac{1}{2}$ we have
\begin{eqnarray}
\label{Pc2}
P_\text{c}(1) &=&  \int_0^\infty dy\, \left[\frac{\partial \mathcal{E}(y)}{\partial y}\,E(0,y)-
\mathcal{E}(y)\,\frac{\partial E(0,y)}{\partial y}\right] \nonumber\\
&-&\frac{1}{2}\int_0^\infty dy\, c(y)
\end{eqnarray}
Integrating by parts, we simplify \eqref{Pc2} to
\begin{equation}
\label{P1:sol}
P_\text{c}(1) = 2\int_0^\infty dy\, \frac{\partial \mathcal{E}(y)}{\partial y}\,E(0,y)-\frac{1}{2}\int_0^\infty dy\, c(y)
\end{equation}
Specifying \eqref{Exyt:sol} to $x=0$, we get
\begin{eqnarray*}
E(0,y) & = & 1 -  \tfrac{1}{2} \text{Erfc}\left(\tfrac{Y}{2}\right)\text{Erf}\left(\tfrac{Y}{2}\right) \\
&-&\frac{1}{\pi}\int_{-\frac{Y}{2}}^\infty du\,e^{-u^2}\int_{u}^{u+Y} dv\,e^{-v^2}
\end{eqnarray*}
Expressing the integral through error functions yields 
\begin{equation*}
E(0,y) =  1 -  \tfrac{1}{2}\text{Erf}\left(\tfrac{Y}{2}\right) +\tfrac{1}{4}\!\left[\text{Erf}\left(\tfrac{Y}{2}\right)\right]^2
-\tfrac{1}{4}\text{Erf}\left(\tfrac{Y}{\sqrt{2}}\right)
\end{equation*}
It is convenient to transform $y$ to $Y$ in the first integral in \eqref{P1:sol}. 
Equation \eqref{EX} gives $dy\, \frac{\partial \mathcal{E}(y)}{\partial y} = dY\,\frac{1}{\sqrt{2\pi}}\,e^{-Y^2/2}$. The above results together with $\int_0^\infty dy\, c(y)=\langle N\rangle_\text{c}$, see \eqref{Nav:int}, allow us to re-write \eqref{P1:sol} as
\begin{equation}
\label{P1:int}
P_\text{c}(1) = \sqrt{\frac{8}{\pi}}\int_0^\infty dz\,e^{-2z^2}\,\Pi(z)-\frac{1}{2}\,\langle N\rangle_\text{c}
\end{equation}
with
\begin{equation*}
\Pi(z) = 1 -  \tfrac{1}{2}\text{Erf}(z) +\tfrac{1}{4}\!\left[\text{Erf}(z)\right]^2
-\tfrac{1}{4}\text{Erf}\big(\sqrt{2}\,z\big)
\end{equation*}
Computing the integrals in \eqref{P1:int} one gets
\begin{equation*}
\begin{split}
& \sqrt{\frac{8}{\pi}}\int_0^\infty dz\,e^{-2z^2} = 1\\
&  \sqrt{\frac{8}{\pi}}\int_0^\infty dz\,e^{-2z^2}\, \text{Erf}(z) = \frac{2}{\pi}\,\arctan\!\left(\frac{1}{\sqrt{2}}\right)\\
& \sqrt{\frac{8}{\pi}}\int_0^\infty dz\,e^{-2z^2}\, \text{Erf}(\sqrt{2}\,z) = \frac{1}{2}\\
&  \sqrt{\frac{8}{\pi}}\int_0^\infty dz\,e^{-2z^2}\, \left[\text{Erf}(z)\right]^2 = \frac{2}{\pi}\,\arctan\!\left(\frac{1}{\sqrt{8}}\right)
\end{split}
\end{equation*}
Using these expressions and recalling  \eqref{Nav:c} we reduce \eqref{P1:int} to the announced result \eqref{CP:one}.

\end{document}